\documentclass[journal]{IEEEtran}
\IEEEoverridecommandlockouts

\newcommand*\Bell{\ensuremath{\boldsymbol\ell}}
\usepackage{cite}
\usepackage{amsmath,amsfonts}

\usepackage{algorithmic}
\usepackage[linesnumbered,ruled,vlined]{algorithm2e}
\usepackage{amssymb}
\usepackage[export]{adjustbox}
\usepackage{graphicx}
\usepackage{mathtools}

\newcommand*{\tran}{^{\mkern-1.5mu\mathsf{T}}}

\newcommand*{\hermconj}{^{\mathsf{H}}}

\usepackage{textcomp}
\usepackage{xcolor}
\usepackage{ragged2e}
\def\BibTeX{{\rm B\kern-.05em{\sc i\kern-.025em b}\kern-.08em
    T\kern-.1667em\lower.7ex\hbox{E}\kern-.125emX}}
\begin{document}

\title{Towards Joint Optimization for UAV-Integrated RIS-Assisted Fluid Antenna Systems}   \author{Ali~Reda,~\IEEEmembership{Student Member,~IEEE,}
Tamer~Mekkawy,~\IEEEmembership{Senior,~IEEE,} Theodoros~A.~Tsiftsis,~\IEEEmembership{Senior, IEEE,}  Chan-Byoung~Chae,~\IEEEmembership{Fellow, IEEE,} and~Kai-Kit~Wong,~\IEEEmembership{Fellow, IEEE}

  \thanks{A. Reda, and T. Mekkawy are with the Avionics Department, Military Technical College, Cairo 11331, Egypt. (e-mails: {ali.aboelyazeed@ieee.org; mekkawy@ieee.org}).}
   \thanks{T. A. Tsiftsis is with the Department of Informatics and Telecommunications, University of the Thessaly, Lamia 35100, Greece, and also with the Department of Electrical and Electronic Engineering, University of Nottingham Ningbo China, Ningbo 315100, China (e-mail: {tsiftsis@uth.gr}).}
   \thanks{ C.-B. Chae is with the School of Integrated Technology, Yonsei University, Seoul 03722, South Korea (e-mail: cbchae@yonsei.ac.kr).}
  \thanks{K.-K.~Wong is with the Department of Electronic and Electrical Engineering, University College London, WC1E 7JE London, U.K., and also with the Yonsei Frontier Laboratory, Yonsei University, Seoul 03722, South Korea (e-mail: kai-kit.wong@ucl.ac.uk).}
  }  
\maketitle
\begin{abstract}
Unmanned aerial vehicles (UAVs) integrated into cellular networks face significant challenges from air-to-ground interference. To address this, we propose a downlink UAV communication system that leverages a fluid antenna system (FAS)-assisted reconfigurable intelligent surface (RIS) to enhance signal quality. By jointly optimizing the FAS port positions and RIS phase shifts, we maximize the achievable rate. The resulting non-convex optimization problem is solved using successive convex approximation (SCA) based on second-order cone programming (SOCP), which reformulates the constraints into a tractable form. Simulation results show that the proposed algorithm significantly improves both outage probability and achievable rate over conventional fixed-position antenna (FPA) schemes, with particularly large gains in large-scale RIS configurations. Moreover, the algorithm converges rapidly, making it suitable for real-time applications.
\end{abstract}


\section{Introduction}
\IEEEPARstart{T}{he} fluid antenna system (FAS) has become a key technology in the field of next-generation wireless communications in recent years. This importance arises from the fact that FAS enables flexible switching of finely positioned elements to the most advantageous port. In contrast, traditional fixed antenna techniques, such as multiple-input multiple-output (MIMO), are constrained by the physical size limitations of wireless devices \cite{P1}. FAS can be practically implemented using liquid metal structures or pixel-based methods \cite{9,10}. Learning-based methods that utilize spatial correlation can help determine the optimal port~\cite{P2}. Related findings have been categorized under the label of movable antenna systems, which can be considered a specific instance of FAS \cite{11}. Fluid antennas (FAs) have attracted considerable attention within the wireless communication field~\cite{1, 23, p4}. In \cite{1}, the authors introduced a FAS featuring a controllable antenna position, representing a unique approach to implementing FAs while enhancing the outage probability analysis. Additionally, \cite{23} explored the sum-rate maximization problem for downlink transmission involving the FA-enabled base station (BS) and multiple users. Furthermore, the authors in~\cite{p4} proposed the integration of MIMO and FAS, demonstrating a significant performance improvement over traditional MIMO systems. Interestingly, FAS has attracted significant interest in current research topics, such as reconfigurable intelligent surface (RIS)~\cite{12,P3}.

Unlike conventional fixed antenna arrays that require multiple RF chains and lack post-deployment flexibility, FAS achieves spatial diversity through dynamic port selection using only a single RF chain \cite{p4}. This reconfigurable capability significantly reduces hardware complexity and power consumption, two critical factors for unmanned aerial vehicles (UAVs) with strict payload and energy constraints. A typical FAS consists of one RF chain and preset locations distributed within a confined surface area \cite{p6}. The radiating element can be rapidly switched among these ports to optimize communication performance, enabling improved achievable rate, lower outage probability, and reduced interference. Since the ports are closely spaced, the resulting channel realizations are strongly correlated, making spatial correlation a central aspect of FAS design \cite{29}. Furthermore, unlike mechanically steerable antennas, FAS adapts instantly to environmental and mobility-induced channel variations without requiring physical movement, making it particularly well-suited for dynamic UAV scenarios.

The fine-grained repositioning capability of FAS allows effective interference mitigation, an advantage that traditional antenna selection (AS) systems with fixed antenna locations cannot provide in practice~\cite{p7}. Additionally, FAS introduces dynamic spatial reconfigurability by allowing the physical position of the radiating element to be adjusted over a predefined surface. This contrasts sharply with traditional AS, where antennas must be spaced at least half a wavelength apart to avoid coupling~\cite{p10}. The ability to relocate a single RF-fed element over a dense grid of virtual ports offers finer spatial resolution and additional degrees of freedom without increasing RF hardware complexity~\cite{p8}. Consequently, FAS achieves superior interference suppression, enhanced beam control, and improved diversity multiplexing trade‑offs under the same RF budget. Furthermore, FAS enables q‑outage capacity improvements and provides a richer optimization space than traditional AS systems, especially in mobile or space‑constrained settings like UAVs~\cite{p9}.

Integrating FAs into communication systems presents a key challenge: the precise and timely adjustment of antenna positions for optimal channel selection. This complex process necessitates sophisticated channel estimation techniques, leading to increased system overhead \cite{13}. Achieving optimal antenna placement requires computationally intensive algorithms, which further complicates system design. To address these challenges, RIS \cite{14} offers a promising solution. By dynamically adjusting the phase of the reflected signal, RIS can effectively modify the wireless channel without requiring mechanical adjustments to the antenna \cite{15}. This approach can simplify the channel adaptation process and reduce the reliance on complex algorithms for optimal antenna positioning.

RIS has emerged as a potential technology that could greatly increase the coverage area in future wireless communication systems. RISs are artificial surfaces that have several inexpensive reflective components that can dynamically adjust the environment in which wireless signals propagate in order to reroute radio waves from BSs to certain mobile users (MUs) \cite{16}. RISs are composed of multiple passive, low-cost elements that can dynamically adjust the phase shift of incident radio waves. Unlike traditional active antennas, RISs do not require RF chains or complex digital signal processing circuits, making them easier to deploy and integrate into various environments. Furthermore, studies \cite{24} have explored the optimization of the sum rate in RISs, highlighting the significant advantages of RIS technology. Finally, \cite{28} explored anti-jamming strategies in RIS-assisted wireless communication to maximize the sum rate under worst-case jamming scenarios.

On the other hand, UAVs have been widely adopted in wireless networks due to their cost-effectiveness, high mobility, and ability to facilitate line-of-sight (LoS) transmission \cite{25}. RIS-assisted UAVs enable intelligent signal reflection in wireless environments. Compared to terrestrial RIS deployments, UAV-based RIS systems are more effective at establishing robust LoS links with ground base stations due to the UAV's elevated position, thereby mitigating signal blockage \cite{26}. Furthermore, in \cite{p5}, the authors proposed a deep reinforcement learning-based optimization algorithm for an RIS-assisted UAV system to enhance covert communication performance by optimizing the UAV's 3D trajectory and the RIS phase shifts.

The integration of FAS and RIS on UAVs addresses key challenges in aerial-ground interference mitigation by providing a powerful yet lightweight solution. FAS enables dynamic spatial diversity within a compact footprint through port switching, eliminating the need for bulky antenna arrays while maintaining aerodynamic efficiency \cite{p12}. RIS complements this by passively enhancing signal quality without active beamforming components \cite{p13}. This integration is particularly advantageous in interference-limited environments, as recent hardware advances in lightweight RIS meta-atoms and microfluidic FAS designs have made such integration feasible for modern UAVs. Together, they preserve UAV agility and energy efficiency while significantly improving performance.

Motivated by the substantial benefits of RIS in FA-assisted communication systems, FAS-RIS systems have emerged as a promising new area of research. In terms of performance analysis, the authors in \cite{16} investigated the performance of RIS-aided FAS systems. Moreover, the authors in \cite{21} proposed a comprehensive framework encompassing performance analysis and throughput optimization. From an optimization perspective, \cite{22} explored low-complexity beamforming design for RIS-assisted FAS systems, while \cite{23} concentrated on sum-rate optimization for RIS-aided multi-user multiple access systems.

Inspired by the preceding discussion, this paper investigates the joint optimization of RIS meta-atom phase shifters and the position of FAs for a novel FA-assisted downlink communication system. In this system, a BS with a fixed-position antennas (FPAs) transmits data to a 1D FA-equipped UAV. We focus on optimizing the system to balance the rate at the UAV and the overall quality of service (QoS). Due to the non-convex nature of the problem with highly coupled variables, we propose a successive convex approximation (SCA) based on the second-order cone programming (SOCP) algorithm to obtain a suboptimal solution. Simulation results demonstrate that the proposed FA-based SCA significantly improves the rate while ensuring QoS through optimal antenna positioning and RIS phase shifters.
The main contributions of this work are summarized as follows:
\begin{itemize}
\item  We introduce an interference mitigation scheme that utilizes a single UAV equipped with an RIS, eliminating the need for deploying multiple RISs near each ground BS \cite{19}. This approach leverages passive beamforming at the RIS to suppress interference from co-channel BSs, effectively mitigating aerial-ground interference during downlink communication. The proposed scheme significantly reduces the complexity and cost associated with traditional multi-RIS deployments while maintaining robust performance.

\item  We formulate a joint optimization problem that simultaneously determines the active fluid antenna positions and RIS phase shifts to maximize the achievable rate at the UAV while mitigating interference from co-channel base stations. Given the non-convex nature of the formulated problem, we develop an SCA-based iterative approach to transform the constraints into a convex SOCP problem, making it computationally efficient and solvable using MOSEK.
    
\item We provide a detailed analysis of how RIS phase shifts optimization and FAS dynamic positioning contribute to improving the signal-to-interference-plus-noise ratio (SINR). Our findings demonstrate that fluid antennas significantly outperform FPA in terms of outage probability and achievable rate, particularly under strong interference conditions.

\item We prove the rapid convergence of the proposed SCA-based algorithm and demonstrate its computational efficiency. By simultaneously updating the FAS positions and RIS phase shifts in each iteration, our approach achieves a stationary point with significantly reduced complexity compared to traditional alternating optimization (AO) methods. This makes the proposed framework highly suitable for practical, real-time UAV communication systems.
\end{itemize}

\begin{figure}[!t]
  \includegraphics[width=0.945\linewidth]{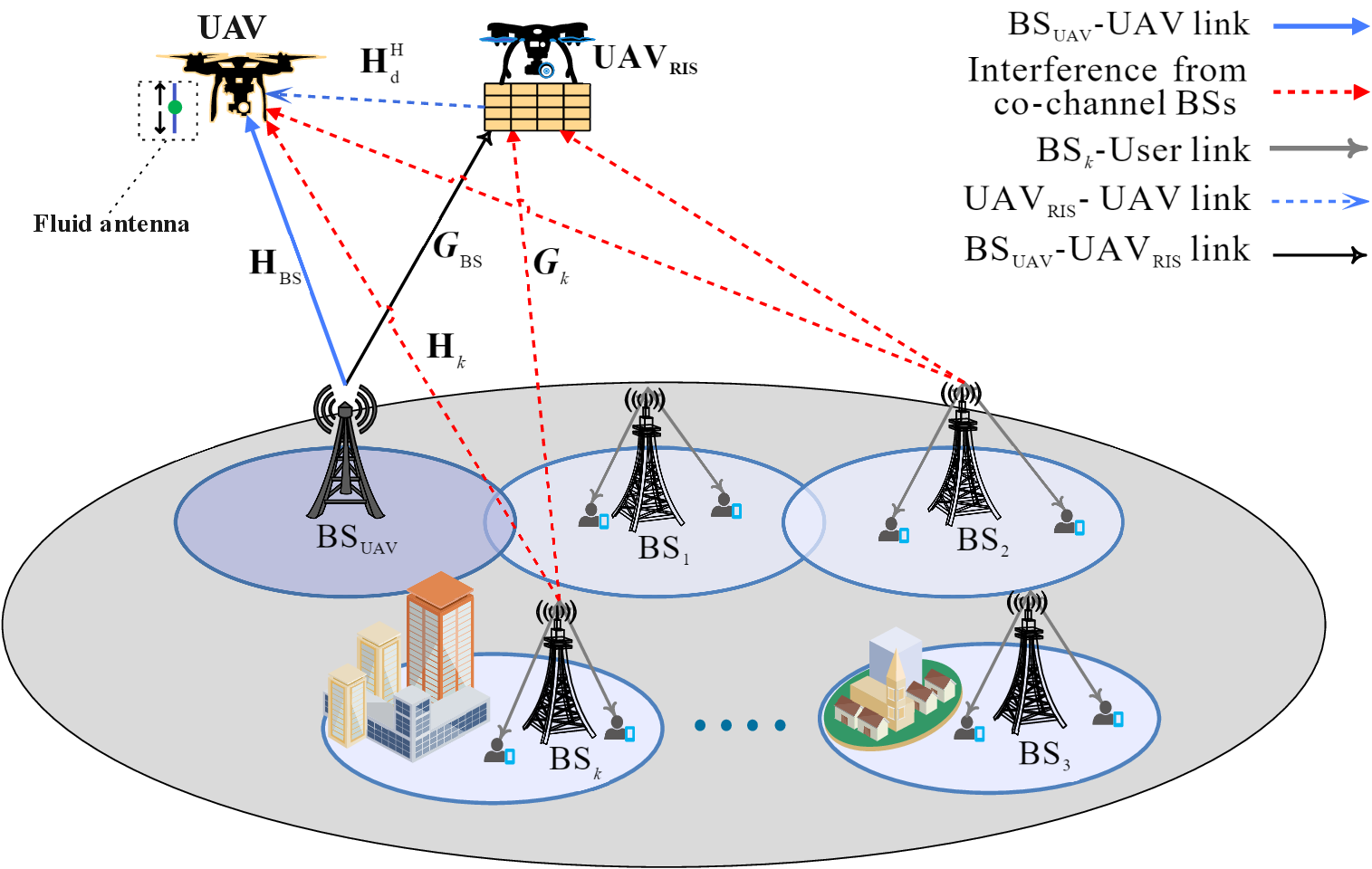}
  \caption{Downlink interference in a cellular-UAV scenario.}
  \label{fig.1}
\end{figure}

\subsubsection*{Notations} Regular and bold small letters stand for scalars and vectors, while bold capital letters are used for matrices. The magnitude of a scalar ${x}$ is represented by $|{x}|$, while the Euclidean norm of a vector ${\textbf{x}}$ is denoted as $\lVert \mathrm{\textbf{x}} \rVert$. The transpose of the matrix $\mathrm{\textbf{X}}$ is represented by $\mathrm{\textbf{X}}\tran$, the conjugate transpose by $\mathbf{X}\hermconj$, and the rank is $\rm{rank}(\mathbf{X})$. The notation $\rm{diag}(\mathbf{X})$ refers to a diagonal matrix. The notation $\mathbb{C}^{m\times n}$ represents the set of $m\times n$ complex matrices. Complex numbers' real and imaginary parts are presented as $\Re(\cdot)$ and $\Im(\cdot)$. Finally, the letter $j$ represent the imaginary unit $\sqrt{-1}$.

\section{System Model} \label{Sec.sys.model}
In this section, we introduce the system model being considered and identify the RIS and FAS modules responsible for maximizing the rate of the UAV.

We consider a far-field downlink cellular scenario, depicted in Fig.~\ref{fig.1}, where a UAV-mounted RIS facilitates communication from a BS to the UAV, mitigating potential channel blockage. The far-field approximation is justified by the assumption that the FAS mobility range is significantly smaller than the transmitter–receiver distance. Consequently, amplitude variations across array elements are negligible, resulting in a simple time-delay relationship between signals. Therefore, the angles of departure (AoDs) and angles of arrival (AoAs) for each propagation path are primarily determined by the scattering environment and remain effectively constant regardless of antenna placement \cite{29}. 

\begin{figure}[!t]
  \includegraphics[width=0.945\linewidth]{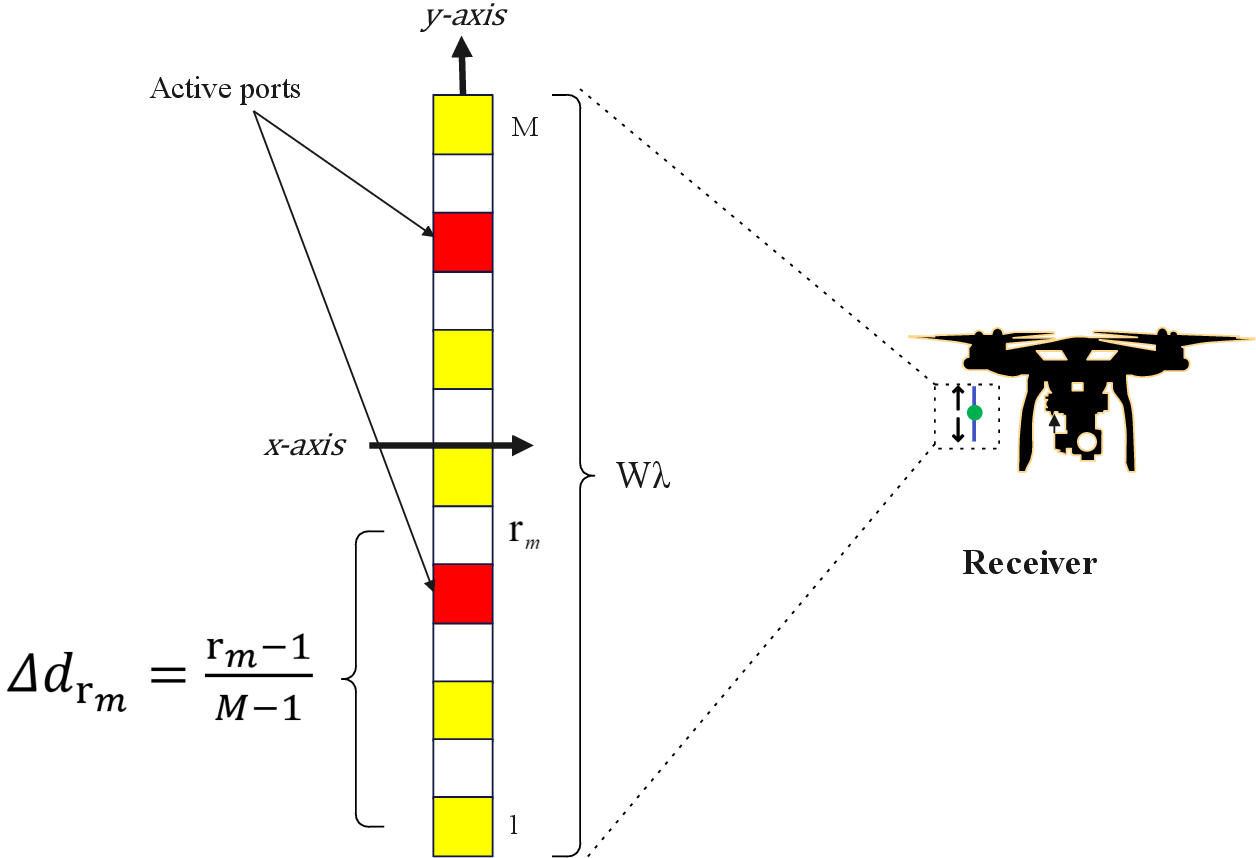}
  \caption{Schematic of a fluid antenna system with $M$ flexible ports.}
  \label{fig.A}
\end{figure}

The UAV communicates with a $\mathrm{BS_{UAV}}$, equipped with $N_t$ antennas, in a scenario with $N$ meta-atoms. The BS is equipped with multiple fixed-position antennas and a 1D FA in UAV for reception. The FA has $M$ flexible positions (or ports), each of which can be activated or deactivated independently along a linear dimension of length, $W \lambda$, where $W$ represents the length of the FA and $\lambda$ denotes the wavelength of radiation, as shown in Fig.~\ref{fig.A}. In this paper, we consider the scenario where multiple ports can be activated simultaneously. This scenario further includes $K$ co-channel BSs serving ground UEs, which may potentially introduce interference for UAV communication. A UAV equipped with an RIS ($UAV_{RIS}$) and a target UAV equipped with FAS are deployed to minimize the interference that is caused in the $k^{th}$ BS, ${\rm{BS}}_k$, $\forall k \in \{1,2,...,K\}$. Here, we assume only aerial-ground interference exists, as shown in Fig.~\ref{fig.1}. Let $\mathrm{UAV_{RIS}}$ have fixed height and elevation angle, and the RIS has $N$ meta-atoms. The local coordinates of the FPA at the $\mathrm{BS_{UAV}}$ are denoted by $[X_{n_t}, Y_{n_t}]$, $1 \leq n_t \leq N_t$, with $O_t$, as the origin. The positions of the reflection elements of the $n$-th RIS element are denoted by $[x_n, y_n]$, $1 \leq n \leq N$.

The meta-atoms in $\mathrm{UAV_{RIS}}$ aim to enhance the power efficiency of the communication link between the $\mathrm{BS_{UAV}}$ and the UAV. Let the phase, $\boldsymbol{\theta}_i \in[0,2\pi)$, $\forall i \in \{1,2,...,N\}$, be adjusted for each meta-atom, with $\alpha_i \left(\boldsymbol{\theta}_i\right)=1$. Consequently, we denote the reflection diagonal matrix associated with the RIS as
\begin{equation}\label{eq.1}
  \boldsymbol{\theta}= \rm{diag} \left[\alpha_1e^{j{\theta}_1} ~~~~ \alpha_2e^{j{\theta}_2} ~....~~~\alpha_\mathrm{N} e^{j{\theta}_\mathrm{N}} \right]\hermconj, \in \mathbb{C}^{\mathrm{N x N}} 
\end{equation}

We establish a local Cartesian coordinate system with the center of UAV's FA, $O_f$, as the origin. The FA is positioned along the $y$-axis, with $M$ ports spaced at intervals of $d_m$. During signal reception, $m_l$ ports are selected or activated. Denote the $m$-th FA’s position as $[x_m,y_m]\tran$, and the indices of the selected port positions are represented by the vector $\boldsymbol{r} = [r_1, \ldots, r_m, \cdots, r_{m_l}]^{\tran} \in \mathbb{R}^{m_l\times 1}$, where each $r_m \in [1, M]$. The coordinates of the activated port $r_m$ are (0, $\mathcal{K}_{r_m} d_{m}$), where $\mathcal{K}$ is given by:
\begin{equation}\label{eq.4}
    \mathcal{K} = \frac{2\left(r_m - 1 \right)-M+1}{2}.
\end{equation}

We designate the number of transmit and receive paths as $L_t$ and $L_r$ respectively. The elevation and azimuth angles at the BS AoDs for the $\upsilon$-th transmit path $(1 \leq \upsilon \leq L_t)$ and the AoAs at the receiving end for the $\kappa$-th path $(1 \leq \kappa \leq L_r)$ are represented as $\vartheta_t^\upsilon$, $\phi_t^\upsilon$, and $\vartheta_t^\kappa$, $\phi_t^\kappa$ respectively. We adopt the planar far-field response model to express the channels. We utilize the planar far-field response model to characterize the channels. In this scenario, altering antenna positions modifies the phases of the path response coefficients but will not affect the AoAs/AoDs or the amplitudes of the path response coefficients for each channel path component \cite{30}. Consequently, the array response of the $\kappa$-th receive path from the RIS to the $m$-th active port of the FA can be expressed as
\begin{equation}
    \mathbf{f}(\mathbf{r}_m) = \begin{bmatrix}
    e^{j\frac{2\pi}{\lambda}\rho_{r}^{L_r}({r}_m)} & \cdots & e^{j\frac{2\pi}{\lambda}\rho_{r}^{L_r}({r}_m)}
\end{bmatrix}^{\tran} \in \mathbb{C}^{L_r \times 1},
\end{equation}
where $\rho_{r}^{L_r}(\mathbf{r}_m) = x_m^r \sin\vartheta_{r,\kappa} \cos\phi_{r,\kappa} + y_m^r \cos \vartheta_{r,\kappa}$ represents the signal propagation difference of the $\kappa$-th path between a single reflecting element and its reception origin $O_f$. Thus, the receive field response matrix of all active ports can be expressed as
\begin{equation}
\mathbf{F}(\mathbf{r}) = \begin{bmatrix}
    \mathbf{f}(\mathbf{r}_1) & \cdots & \mathbf{f}(\mathbf{r}_{m_l})
\end{bmatrix} \in \mathbb{C}^{L_r \times m_l},
\end{equation}

Similarly, the array response of the $\upsilon$-th transmit path from the transmit antennas $\mathrm{BS_{UAV}}$ can be expressed as
\begin{equation}
    \mathbf{\zeta}(\mathbf{t}_{n_t}) = \begin{bmatrix}
    e^{j\frac{2\pi}{\lambda}\rho_{t}^{L_t}(\mathbf{t}_{n_t})} & \cdots & e^{j\frac{2\pi}{\lambda}\rho_{t}^{L_t}(\mathbf{t}_{n_t})}
\end{bmatrix}^{\tran} \in \mathbb{C}^{L_t \times 1},
\end{equation}
where $\rho_{t}^{L_t}(\mathbf{t}_{n_t}) = X_{n_t}^t \sin\vartheta_{t,\upsilon} \cos\phi_{t,\upsilon} + Y_{n_t}^t \cos \vartheta_{t,\upsilon}$ represents the signal propagation difference of the position of the $n_t$-th transmit antenna and origin $O_t$ in the $\upsilon$-th transmit path. Thus, the transmit field response matrix at the transmitter can be expressed as
\begin{equation}
    \mathbf{\Lambda}(\mathbf{t}) = \begin{bmatrix}
    \mathbf{\zeta}(\mathbf{t}_{1}) & \cdots & \mathbf{\zeta}(\mathbf{t}_{N_t})
\end{bmatrix} \in \mathbb{C}^{L_t \times N_t}.
\end{equation}

Again, we have considered the phase differences across all transmit paths between the $n$-th RIS reflection element and the UAV, additionally receive paths from the BS to the $n$-th RIS reflection element.

We define the path response matrix $\boldsymbol{\Psi} \in \mathbb{C}^{L_r \times L_t}$ to represent the combined responses of all transmit and receive paths between the transmitter and the receiver. Subsequently, we illustrate the channels connecting the BS with the RIS, the BS with the target UAV, and the RIS with the UAV with the subscripts BR, BU, and RU, respectively. Let $\mathrm{\textbf{H}}_{\mathrm{BS}} = \mathbf{F}\hermconj(\mathbf{r})_{\mathrm{BU}} \boldsymbol{\Psi}_{\mathrm{BU}} \mathbf{\Lambda}(\mathbf{t})_{\mathrm{BU}} \in\mathbb{C}^{m_l \times N_t}$ and $\mathrm{\textbf{H}}_k\in\mathbb{C}^{m_l \times N_t}$ are the direct channel from $\mathrm{BS_{UAV}}$ to the active ports of FA at the UAV and the interfering channel from $\mathrm{BS}_k$ to UAV, respectively, while the channel from RIS to the UAV is ${\mathrm{\textbf{H}}}_{\mathrm{d}}\hermconj = \mathbf{F}\hermconj_{\mathrm{RU}}(\mathbf{r}) \boldsymbol{\Psi}_{\mathrm{RU}} \mathbf{\Lambda}_{\mathrm{RU}}(\mathbf{t}) \in \mathbb{C}^{m_l\times N}$, where $\mathbf{\Lambda}_{\mathrm{RU}}(\mathbf{t}) \in \mathbb{C}^{L_t \times N}$ represents the transmit field response matrix at the RIS. $\mathrm{\textbf{G}_{BS}}$ and $\mathrm{\textbf{G}}_{k}$, represent the channels from $\mathrm{BS_{UAV}}$ and $\mathrm{BS}_k$ to $\mathrm{UAV_{RIS}}$, respectively. Of note, $\mathrm{\textbf{G}_{BS}}$ and $\mathrm{\textbf{G}}_{k}$ $\in \mathbb{C}^{N\times N_t}$.
Given that the target UAV and the $\mathrm{UAV_{RIS}}$ are relatively close to each other but both are located farther from the BS, the BS–UAV and BS–$\mathrm{UAV_{RIS}}$ links are characterized by a combination of strong LoS and scattered non-LoS (NLoS) components, due to the high altitude of the UAVs. Therefore, these links are modeled using Rician fading channels. A generic Rician channel matrix $\mathbf{Z}$ (which represents the channel links, $\mathbf{H}_{BS}$, $\mathbf{G}_{BS}$, $\mathbf{H}_{k}$, or $\mathbf{G}_{k}$) is expressed as:

\begin{equation}
    {\mathrm{\textbf{Z}}} = \sqrt{\frac{\bar{K}}{\bar{K}+1}} {\mathrm{\textbf{Z}}}_{\text{LoS}} + \sqrt{\frac{1}{\bar{K}+1}} {\mathrm{\textbf{Z}}}_{\text{NLoS}}
\end{equation}
\noindent where ${\mathrm{\textbf{Z}}}_{\text{LoS}}$ represents the deterministic LoS component (typically dependent on the path geometry), and ${\mathrm{\textbf{Z}}}_{\text{NLoS}} \sim \mathcal{CN}(0, \mathbf{I})$ denotes the random scattered NLoS component. The Rician factor $\bar{K}$ determines the relative strength of the LoS path. We denote the transmit power vectors of $\mathrm{BS}_{k}$ and $\mathrm{BS_{UAV}}$ as $\textbf{p}_k$ and $\textbf{p}_{\text{BS}}=[p_{1},...,p_{N_{t}}]^{\tran}\in\mathbb{C}^{N_{t}\times1}$, which is a fixed, known complex transmit vector satisfying the total power budget $P_{\max}$. The received signal at the activated ports $m_l$ in the fluid antenna of the UAV can be represented as:

\begin{equation} \label{eq.2}
\begin{aligned}
    \textbf{y}_{m_l} &=\left(\mathrm{\textbf{H}}_{\mathrm{BS}}+{\mathrm{\textbf{H}}}_{\mathrm{d}}\hermconj \boldsymbol{\theta} \mathrm{\textbf{G}_{BS}}\right)  {\textbf{p}_{\text{BS}}}  \ x_{\mathrm{BS}} \\ 
    & \quad +\sum_{k \in \mathcal{S}}{\left(\mathrm{\textbf{H}}_k+{\mathrm{\textbf{H}}}_{\mathrm{d}}\hermconj \boldsymbol{\theta} \mathrm{\textbf{G}}_{k}\right) {\textbf{p}_k}  \ x_k} + \textbf{n}_{m_l}, 
    \end{aligned}
\end{equation}
where the symbols transmitted by $\mathrm{BS_{UAV}}$ and $\mathrm{BS}_k$ are respectively denoted as $x_\mathrm{BS}$, $x_k$, and $\textbf{n}_{m_l} \in \mathbb{C}^{m_l\times 1} \sim\mathcal{CN}\left(0,\sigma_{u}^2 \mathbf{I_{m_l}}\right)$ is the complex additive white Gaussian noise (AWGN) with zero-mean at the $m$ receive ports and variance $\sigma_u^2$ at the UAV. Therefore, the SINR at the target UAV is given by:
\begin{equation}\label{eq.3}
    \beta =\frac{\|\Bell_{\text{BS}}  {\textbf{p}_{\text{BS}}}\|^2 }{\sum_{k \in \mathcal{S}}{\|{\Bell_{k} {\textbf{p}_k}\|^2} }+ m_l\sigma_u^2}
\end{equation}
where,\begin{align*}
    \Bell_{\text{BS}} &=\left(\mathrm{\textbf{H}}_{\mathrm{BS}}+{\mathrm{\textbf{H}}}_{\mathrm{d}}\hermconj \boldsymbol{\theta} \mathrm{\textbf{G}_{BS}}\right), \in \mathbb{C}^{m_l \times N_t} \\
    \Bell_{k} &= \left(\mathrm{\textbf{H}}_k+{\mathrm{\textbf{H}}}_{\mathrm{d}}\hermconj \boldsymbol{\theta} \mathrm{\textbf{G}}_{k}\right), \in \mathbb{C}^{m_l \times N_t} 
\end{align*} 

Evidently, the interference stemming from co-channels is directly influenced by the arrangement of RIS meta-atoms. Hence, the selection of $\alpha_i$ and $\boldsymbol{\theta}_i$ for individual RIS elements plays a crucial role in optimizing $\beta$. In this paper, we assume that the perfect instantaneous channel state information (CSI), specifically the LoS component, is available at both the transmitter and the RIS. Consequently, the achievable rate $R$ (in bps/Hz) for a Gaussian source can be expressed as follows:
\begin{equation}
  R (\boldsymbol{\theta}, \boldsymbol{r}) = \log_2\left(1+\frac{\|\Bell_{\text{BS}} \ \textbf{p}_{\text{BS}}\|^2 }{\sum_{k \in \mathcal{S}}{\|{\Bell_{k} \textbf{p}_k \|^2} }+m_l \sigma_u^2}\right)
\end{equation} 

While RIS elements can sometimes act as active components for channel estimation, this paper considers only passive elements and assumes perfect CSI between the UAV and each BS \cite{19}. This assumption is justified by the strong LoS links typically present between an elevated UAV and ground terminals.
\section{Problem Definition} \label{sec:prob_def}

In this paper, we aim to maximize the UAV's rate while meeting the SINR requirements of the UAV by jointly optimizing the FAS's antenna positions, and the phase shifts of the RIS's meta-atoms. We assume perfect CSI for the communication channels at both the BS and UAV. The corresponding optimization problem is then formulated as follows:
\begin{subequations}\label{eq.7}
\begin{align}
\max_{\boldsymbol{\theta}, \boldsymbol{r}} \quad & R(\boldsymbol{\theta}, \boldsymbol{r}) \\
\textrm{s.t.} \quad & \frac{\|\Bell_{\text{BS}}  \textbf{p}_{\text{BS}}\|^2 }{\sum_{k \in \mathcal{S}}{\|{\Bell_{k} \textbf{p}_k \|^2} }+ m_l \sigma_u^2} \geq \gamma\\
& \boldsymbol{r} = [r_1, \ldots, r_m, \cdots, r_{m_l}]^{\tran} \in \mathbb{R}^{m_l\times 1},\\
& r_m \in [1, M],\\
& r_1 < r_2 < \ldots < r_{m_l}, \\
& \boldsymbol{\theta}_i \in [0,2\pi),\forall i \in \mathcal{N}
\end{align}
\end{subequations}

Constraints (\ref{eq.7}C), (\ref{eq.7}d), and (\ref{eq.7}e) specify that the indices of the activated ports are positive integers between 1 and $M$, arranged in ascending order. The optimization problem is non-convex due to the coupling between the variables $\boldsymbol{r}$ and $\boldsymbol{\theta}$ in the SINR expression. This makes it challenging to solve using traditional convex optimization techniques.

Although several efficient methods exist for solving the problems following AO, it is a non-convex quadratically constrained quadratic program, making optimal solutions challenging to achieve. The authors of \cite{6} employed the semidefinite relaxation (SDR) method with Gaussian randomization. However, this approach has two main drawbacks. First, the complexity of SDR escalates rapidly with the number of reflecting elements, as the problem is transformed into the positive semidefinite domain. Second, obtaining a rank-1 feasible solution necessitates a substantial number of randomization steps, significantly increasing overall complexity. Moreover, AO-based methods are often inefficient because theoretical guarantees of convergence to a stationary point are not available.
\section{Proposed Algorithm}
In this section, we address the significant complexities presented in (\ref{eq.7}) by utilizing a provably convergent SCA-based approach within a SOCP framework. We streamline the optimization process by simultaneously updating the positions of the FAS ($\mathbf{r}$) and the phase shifts of the RIS ($\boldsymbol{\theta}$) in each iteration, while the transmit power vector $\mathbf{p}_{BS}$ is treated as a fixed parameter.

First, to tackle the non-convexity in (\ref{eq.7}), we employ the SCA method to derive a high-performance solution. Our approach utilizes a set of (in)equalities for arbitrary complex-valued $\textbf{a}$ and $\textbf{b}$, as detailed in \cite{4}:
\begin{subequations}\label{eq.8}
\begin{align}
	&\|\textbf{a}\|^2 \geq 2 \Re\{\textbf{b}^{\hermconj} \textbf{a}\}-\|\textbf{b}\|^2,\\
	& \Re\{\textbf{a}^{\hermconj} \textbf{b}\} = \frac{1}{4} \left(\|\textbf{a}+\textbf{b}\|^2 - \|\textbf{a}-\textbf{b}\|^2 \right),\\
	& \Im \{\textbf{a}^{\hermconj} \textbf{b}\} = \frac{1}{4} \left(\|\textbf{a}-j\textbf{b}\|^2 - \|\textbf{a}+j\textbf{b}\|^2 \right).
\end{align}
\end{subequations}

The term in (\ref{eq.7}a) is therefore identified as neither convex nor concave. The non-convexity arises from the complex dependence of the channel vectors $\Bell_{\text{BS}}$ and $\Bell_{\text{k}}$ on the optimization variables, $\mathbf{r}$ and $\boldsymbol{\theta}$. To maximize the function in (\ref{eq.7}a), we apply the SCA method to derive a concave lower bound for the signal power term $\mathbf{\|\Bell_{\text{BS}}\mathbf{p}_{BS}\|^2}$. We treat the entire complex signal vector $\mathbf{a} = \Bell_{\text{BS}}(\mathbf{r}, \theta)\mathbf{p}_{BS}$ as the function to be linearized. We use the inequality (\ref{eq.8}a) by setting the fixed point $\mathbf{b} = \mathbf{u}^{(n)} =\Bell_{\text{BS}}^{(n)}\mathbf{p}_{BS}$, which is the function's value at the $n^{th}$ iteration:
\begin{equation}\label{eq.9}
\begin{aligned}
	\|\Bell_{\text{BS}} \textbf{p}_{\text{BS}}\|^2 &\overset{(a)}{\geq} 2\Re\{\big({\text{u}}^{\left(n\right)}\big)^{\hermconj} \Bell_{\text{BS}} {\textbf{p}_{\text{BS}}}\} - \|{\textbf{u}}^{\left(n\right)}\|^2\\
	&\overset{(b)}{\geq} \frac{1}{2}\{\|{{\text{u}}^{\left(n\right)} \Bell_{\text{BS}}^{\hermconj}+ {\textbf{p}_{\text{BS}}}\|}^2  - \|{{\text{u}}^{\left(n\right)} \Bell_{\text{BS}}^{\hermconj}- {\textbf{p}_{\text{BS}}}\|}^2 \}\\
    &\quad - \|{\textbf{u}}^{\left(n\right)}\|^2 \\
	&\overset{(c)}{\geq} [\,\Re\{\big({\textbf{v}}^{\left(n\right)}\big)^{\hermconj}[\, {\text{u}}^{\left(n\right)} \Bell_{\text{BS}}^{\hermconj} -{\textbf{p}_{\text{BS}}}]\,\} - \frac{1}{2}\|{{\textbf{v}}^{\left(n\right)}\|}^2\\
	&\quad -\frac{1}{2} \|{ {\text{u}}^{\left(n\right)} \Bell_{\text{BS}}^{\hermconj} -{\textbf{p}_{\text{BS}}}\|}^2 - \|{\textbf{u}}^{\left(n\right)}\|^2]\,\\
	& = f\big({\boldsymbol{r}}; \boldsymbol{\theta}; {\boldsymbol{r}^{\left(n\right)}};\boldsymbol{\theta}^{\left(n\right)}\big),
\end{aligned}
\end{equation}
where ${\textbf{u}}^{\left(n\right)} = \Bell_{\text{BS}}^{\left(n\right)} {\textbf{p}_{\text{BS}}}$, ${\textbf{v}}^{\left(n\right)} = {\text{u}}^{\left(n\right)}\big(\Bell_{\text{BS}}^{\left(n\right)}\big)^{\hermconj} + {\textbf{p}_{\text{BS}}}$, and $\Bell_{\text{BS}}^{\left(n\right)} = \mathrm{\textbf{H}}_{\mathrm{BS}}+{\mathrm{\textbf{H}}}_{\mathrm{d}}\hermconj {\boldsymbol{\theta}}^{\left(n\right)} \mathrm{\textbf{H}}_{\mathrm{BS}}$. Here, ${\boldsymbol{\theta}}^{\left(n\right)}$ and and $\Bell_{\text{BS}}^{\left(n\right)}$ denotes the value of $\Bell_{\text{BS}}$ in the $n^{th}$ iteration of the SCA model. Additionally, $(a)$ and $(c)$ in (\ref{eq.9}) result from (\ref{eq.8}a), while $(b)$ results from (\ref{eq.8}b). It can be proven that $f\big({\boldsymbol{r}}; \boldsymbol{\theta}; {\boldsymbol{r}^{\left(n\right)}};\boldsymbol{\theta}^{\left(n\right)}\big)$ is jointly concave w.r.t ${\boldsymbol{\theta}}$ and ${\boldsymbol{r}}$.

Next, to tackle (\ref{eq.7}b) non-convex SINR constraint, we apply the SCA method, optimizing both $\boldsymbol{\theta}$ and ${\boldsymbol{r}}$ during each iteration. Initially, we convert (\ref{eq.7}b) into an equivalent expression as follows:
\begin{subequations}\label{eq.10}
\begin{align}
	&\frac{\|\Bell_{\text{BS}} \textbf{p}_{\text{BS}}\|^2 } {\gamma} \geq {m_l \sigma_u}^2 + \sum_{k \in \mathcal{S}}{\big({{\varrho}_k}^2+{{\bar{\varrho}}_k}^2\big),} \\
	&\quad \quad {\varrho}_k \geq \|\Re{\{{\Bell_{k} \textbf{p}_k \}}}\|, \forall k \in \mathcal{S}, \\
	&\quad \quad {\bar{\varrho}}_k \geq \|\Im{\{{\Bell_{k} \textbf{p}_k\}}}\|, \forall k \in \mathcal{S}
\end{align}      
\end{subequations}
where the new slack variables ${\varrho}_k$ and ${\bar{\varrho}}_k$ are used. It is simple to observe that if (\ref{eq.7}b) is feasible, then (\ref{eq.10}) is also feasible, and vice versa. Since the right-hand side of (\ref{eq.10}a) is convex, we need to find a concave lower bound for the term $\|\Bell_{\text{BS}} \textbf{p}_{\text{BS}}\|^2$ in (\ref{eq.10}a). Let ${\boldsymbol{r}^{\left(n\right)}}$ and $\boldsymbol{\theta}^{\left(n\right)}$ represent the value of $\boldsymbol{r}$ and $\boldsymbol{\theta}$ in the $n^{th}$ iteration of the SCA process, respectively. Similarly to (\ref{eq.9}), this can be represented as:
\begin{equation}\label{eq.11}
\begin{aligned}
\frac{\|\Bell_{\text{BS}} \textbf{p}_{\text{BS}}\|^2 } {\gamma} &\geq  \frac{1} {\gamma}  [\,\Re\{\big({\textbf{v}}^{\left(n\right)}\big)^{\hermconj}[\, {\text{u}}^{\left(n\right)} \Bell_{\text{BS}}^{\hermconj} -{\textbf{p}_{\text{BS}}}]\,\} \\
  -\frac{1}{2}\|{{\textbf{v}}^{\left(n\right)}\|}^2
	& -\frac{1}{2} \|{ {\text{u}}^{\left(n\right)} \Bell_{\text{BS}}^{\hermconj} -{\textbf{p}_{\text{BS}}}\|}^2 - \|{\textbf{u}}^{\left(n\right)}\|^2] \\
	& = f\big({\boldsymbol{r}}; \boldsymbol{\theta}; {\boldsymbol{r}^{\left(n\right)}}; \boldsymbol{\theta}^{\left(n\right)}\big).
\end{aligned}
\end{equation}

By utilizing the property that $x \geq |y|$ if and only if $x \geq y$ and $x \geq -y$, and incorporating (\ref{eq.8}b), we can express (\ref{eq.10}b) in an equivalent form as
\begin{subequations}\label{eq.12}
\begin{align}
	&{\varrho}_k \geq \Re{\{{\Bell_{k} {\textbf{p}_k}\}}} = \frac{1}{4} \big(\|\Bell_{k}^{\hermconj}+{\textbf{p}_k}\|^2 - \|\Bell_{k}^{\hermconj}-{\textbf{p}_k}\|^2  \big), \\
	&{\varrho}_k \geq -\Re{\{{\Bell_{k} {\textbf{p}_k}\}}} = \frac{1}{4} \big(\|\Bell_{k}^{\hermconj}-{\textbf{p}_k}\|^2 - \|\Bell_{k}^{\hermconj}+{\textbf{p}_k}\|^2  \big).
\end{align}      
\end{subequations}

The negative quadratic term on the right-hand side of (\ref{eq.12}a) renders it non-convex. Consequently, we can reformulate it into a convex form by applying the inequality from (\ref{eq.8}a) as follows:
\begin{equation}\label{eq.13}
\begin{aligned}
	{\varrho}_k &\geq \frac{1}{4} [\|\Bell_{k}^{\hermconj}+{\textbf{p}_k}\|^2 - 2\Re\{\big(\Bell_{k}^{\left(n\right)} - \big({\textbf{p}_k}\big)^{\hermconj} \\
 &\big(\Bell_{k}^{\hermconj}-{\textbf{p}_k}\big)\} + \|\big({\Bell_{k}}^{\left(n\right)}\big)^{\hermconj}-{\textbf{p}_k}\|^2 ] \\
 & \quad = \rho_k({\boldsymbol{r}}; \boldsymbol{\theta}; {\boldsymbol{r}^{\left(n\right)}};\boldsymbol{\theta}^{\left(n\right)}\big).      
\end{aligned}
\end{equation}
Using a similar reasoning, (\ref{eq.12}b) leads to:
\begin{equation}\label{eq.14}
\begin{aligned}
	{\varrho}_k &\geq \frac{1}{4} [\|\Bell_{k}^{\hermconj}-{\textbf{p}_k}\|^2 - 2\Re\{ \big(\Bell_{k}^{\left(n\right)} + \big({\textbf{p}_k}\big)^{\hermconj}\big) \\
 &\big(\Bell_{k}^{\hermconj}+{\textbf{p}_k}\big)\} + \|\big({\Bell_{k}}^{\left(n\right)}\big)^{\hermconj}+{\textbf{p}_k}\|^2 ] \\
 & \quad = \bar{\rho}_k({\boldsymbol{r}}; \boldsymbol{\theta}; {\boldsymbol{r}^{\left(n\right)}};\boldsymbol{\theta}^{\left(n\right)}\big).      
\end{aligned}
\end{equation}
Similar (\ref{eq.13}) and (\ref{eq.14}), (\ref{eq.10}c) results in the following inequalities:
\begin{equation}\label{eq.15}
\begin{aligned}
	{\bar{\varrho}}_k &\geq \frac{1}{4} [\|\Bell_{k}^{\hermconj}-j{\textbf{p}_k}\|^2 - 2\Re\{\big(\Bell_{k}^{\left(n\right)} - j\big({\textbf{p}_k}\big)^{\hermconj} \\
 & \big(\Bell_{k}^{\hermconj}+j{\textbf{p}_k}\big)\} + \|\big({\Bell_{k}}^{\left(n\right)}\big)\hermconj+j{\textbf{p}_k}\|^2 ] \\
 & \quad = \omega_k({\boldsymbol{r}}; \boldsymbol{\theta}; {\boldsymbol{r}^{\left(n\right)}};\boldsymbol{\theta}^{\left(n\right)}\big).
\end{aligned}
\end{equation}

\begin{equation}\label{eq.16}
\begin{aligned}
	{\bar{\varrho}}_k &\geq \frac{1}{4} [\|\Bell_{k}^{\hermconj}+j{\textbf{p}_k}\|^2 - 2\Re\{ \big(\Bell_{k}^{\left(n\right)} + j\big({\textbf{p}_k}\big)^{\hermconj}\big) \\
 & \big(\Bell_{k}^{\hermconj}-j{\textbf{p}_k}\big)\} + \|\big({\Bell_{k}}^{\left(n\right)}\big)^{\hermconj}-j{\textbf{p}_k}\|^2 ] \\
 &\quad = \bar{\omega}_k({\boldsymbol{r}}; \boldsymbol{\theta}; {\boldsymbol{r}^{\left(n\right)}};\boldsymbol{\theta}^{\left(n\right)}\big).      
\end{aligned}
\end{equation}
Subsequently, the non-convexity of (\ref{eq.7}e) is the only remaining challenge. We add a regularization term to the goal function to guarantee that the inequality requirement is satisfied as an equality at convergence. Furthermore, we use a first-order approximation of the regularization term centered at $\boldsymbol{\theta}^{\left(n\right)}$ to address the non-convexity of the resulting objective function. To obtain an equivalent convex sub-problem of the SCA model, we reformulate the non-convex problem in  (\ref{eq.7}) as:

\begin{subequations}\label{eq.17}
\begin{align}
\max_{\boldsymbol{\theta}, \boldsymbol{r}, \tau, \Bar{\tau}} \quad & \beta(\boldsymbol{\theta}, \boldsymbol{r}) + {\delta[2\Re{ \{\big(\boldsymbol{\theta}^{\left(n\right)}\big)\hermconj \boldsymbol{\theta}\}}- \lVert \boldsymbol{\theta}^{\left(n\right)} \rVert^2}], \\
\textrm{s.t.} \quad & f({\boldsymbol{r}}; \boldsymbol{\theta}; {\boldsymbol{r}^{\left(n\right)}};\boldsymbol{\theta}^{\left(n\right)}\big) \geq {m_l \sigma_u}^2 \\ &  + \sum_{k \in \mathcal{S}}{\big({{\varrho}_k}^2+{{\bar{\varrho}}_k}^2\big),} \quad \forall k \in \mathcal{S}\\
& \quad (\ref{eq.13})-(\ref{eq.16}), \quad \forall k \in \mathcal{S}\\
& \boldsymbol{r} = [r_1, \ldots, r_m, \cdots, r_{m_l}]^{\tran} \in \mathbb{R}^{m_l\times 1},\\
& r_m \in [1, M],\\
& r_1 < r_2 < \ldots < r_{m_l},\\
& \boldsymbol{\theta}_i \in [0,2\pi),\forall i \in \mathcal{N}   
\end{align}
\end{subequations}
where $\tau = \{\varrho_k\}_{k \in \mathcal{S}}$, $\Bar{\tau} = \{\bar{\varrho}_k\}_{k \in \mathcal{S}}$, $n$ denotes the iteration number, and $\delta > 0$ is the regularization parameter. The non-convexity of (\ref{eq.17}) arises from the regularization term involving $\delta$ in (\ref{eq.17}a). To address this, we employ (\ref{eq.8}a) to convexify (\ref{eq.17}a). Each constraint in (\ref{eq.17}) can be expressed as a second-order cone constraint, making it a SOCP problem that can be efficiently solved using the MOSEK solver \cite{5}. The optimization process iteratively updates the RIS phase shifts $\boldsymbol{\theta}$ and the FAS positions $\boldsymbol{r}$. When the objective function converges to a value within a predefined tolerance $\epsilon$, the optimal solutions $\boldsymbol{\theta}^{\star}$ and $\boldsymbol{r}^{\star}$ are obtained through eigenvalue decomposition.

The integration of UAV-mounted FAS and RIS technologies, although seemingly complex for a single-UAV setup, addresses critical limitations of static, ground-based architectures. By leveraging the mobility of UAVs and the spatial adaptability of FAS, the system can dynamically suppress interference, significantly outperforming traditional fixed arrays. Moreover, the RIS component can emulate the coverage of multiple ground units, reducing infrastructure costs. While our current evaluation considers a single UAV, the framework serves as a scalable foundation for future multi-UAV deployments, enabling distributed communication, sensing, and interference management through lightweight, reconfigurable hardware.

\section{Algorithm Analysis}\label{sec.4}
The proposed algorithm, outlined in Algorithm \ref{algorithm1}, focuses on optimizing $\boldsymbol{\theta}$ to maximize the rate by adjusting $\epsilon$ through the MOSEK solver. This section delves into outage probability analysis and the feasibility of interference elimination in LoS channels, followed by discussions on convergence behavior and computational complexity.

\subsection{Outage Probability Analysis}

The outage probability IS defined as the probability that the target SINR of the UAV falls below a certain threshold $\gamma$. This threshold ensures that the communication link meets the required QoS. The outage event occurs when the SINR at the UAV is less than the target SINR threshold $\gamma$. Let $R$ denote a specific target transmission rate, the outage probability can be formulated as
\begin{equation}
P_{\text{out}} = P_r\left( \log_2\left(1+\beta\right) < \gamma \right) =F_\beta \left(\gamma \right).
\end{equation}
where $F_\beta \left(\gamma \right)=2^R-1$. Obtaining the cumulative distribution function (CDF) of $\beta$, $F_\beta \left(\cdot \right)$ is crucial to assessing outage probability. By leveraging the central limit theorem (CLT), we can analyze the distribution of $\beta$ assuming a large number of RIS meta-atoms \cite{12}. The numerator can be approximated as a Gamma-distributed random variable due to the squared magnitude of a complex Gaussian. Then, the denominator of $\beta$ is a sum of independent random variables (the interference plus noise), which can also be approximated as a Gaussian random variable due to the CLT. In this paper, we have determined the outage probability for the optimal solutions of Algorithm \ref{algorithm1}, denoted as $\boldsymbol{\theta}^{\star}$ and $\boldsymbol{r}^{\star}$, after the objective function in Equation (\ref{eq.17}) has converged.

\subsection{Feasibility of Interference Elimination in LoS Channels}
In this paper, we explore a specific scenario involving exclusively LoS channels connecting the BS-UAV, $\mathrm{UAV_{RIS}}$-UAV, and BS-$\mathrm{UAV_{RIS}}$, aiming to analyze how the quantity of reflecting elements influences the feasibility of interference cancellation. This scenario holds practical importance due to the heightened altitudes of the UAV and UAV-RIS. As a result, we can assess the channel gains for both the direct and cascaded paths associated with $\text{BS}_k$.
\begin{subequations}\label{eq.18}
\begin{align} 
    &\|{\mathrm{\textbf{H}}_k}\|_1 = \frac{\sqrt{\beta_0}}{d_{\text{BU},k}}, k \in \mathcal{S} \\
    &\|{\mathrm{\textbf{H}}_{\mathrm{d}}^\mathrm{H} \boldsymbol{\theta} \mathrm{\textbf{G}}_{k}}\|_1 = \frac{N \beta_0}{d_{\text{BI},k} d_{\text{IU}}}, k \in \mathcal{S}
\end{align}
\end{subequations}

In this context, $\beta_0$ represents the LoS path loss at a reference distance of 1 meter. The variables $d_{\text{BU},k}$, $d_{\text{IU}}$, and $d_{\text{BI},k}$ denote the distances between the $\text{BS}_k$-UAV, $\mathrm{UAV_{RIS}}$-UAV, and $\text{BS}_k$-$\mathrm{UAV_{RIS}}$ links, respectively. By substituting (\ref{eq.18}a) and (\ref{eq.18}b) into the complete interference nulling condition, $\|{\mathbf{H}_\text{d}^\text{H} \boldsymbol{\theta} \mathbf{G}_k}\|_1 \geq \|{\mathrm{\textbf{H}}_k}\|_1$, we obtain:

  \begin{equation} \label{eq.19}
     \frac {N \beta_0} {d_{\text{BI},k} d_{\text{IU}}} \geq \frac{\sqrt{\beta_0}}{d_{\text{BU},k}}, k \in \mathcal{S} 
  \end{equation}     

In this scenario, the similar altitudes of the UAV and $\mathrm{UAV_{RIS}}$ lead to approximately equal distances between the $\text{BS}_k$-UAV and $\text{BS}_k$-$\mathrm{UAV_{RIS}}$ links, \textit{i.e.}, $d_{\text{BU},k} \approx d_{\text{BI},k}$. Under these conditions, the minimum number of reflecting elements needed to achieve interference nulling at the UAV can be approximated as $N_{min} = \left\lceil\frac{d_{{\text{IU}}}}{\sqrt{\beta_0}}\right\rceil$. As shown through simulations in Section~\ref{sec.5}, a larger number of reflecting elements $N$ facilitates better interference nulling performance.

\begin{algorithm}[t!] 
	\caption{SCA-based Joint Optimization Algorithm for (\ref{eq.17})}\label{algorithm1}
	\SetAlgoLined
	\LinesNumbered
	\SetKwInOut{Input}{Input}
	\Input{Maximum iteration number $N_m$, initialization $\boldsymbol{\theta}^{(0)}, \boldsymbol{r}^{(0)}$, tolerance $\delta>0$, step size $\eta$}
	\SetKwInOut{Output}{Output}
	\Output {Optimized variables $\{{\boldsymbol{\theta}}^{\star},\boldsymbol{r}^{\star }\}$.}
	\textbf{Initialize}: $n=0$; $\Delta R = \infty$;\\
	\Repeat{$\Delta R < \delta$ or $n \geq N_m$}{
		\textbf{Solve} (\ref{eq.17}) via MOSEK to obtain $\boldsymbol{\theta}^{(n+1)}, \boldsymbol{r}^{(n+1)}$; \\
		\textbf{Project} $\boldsymbol{\theta}^{(n+1)}$ onto $[0,2\pi)$ and $\boldsymbol{r}^{(n+1)}$ onto feasible FAS positions;\\
		\textbf{Update} variables using a relaxation step: 
		\begin{align*}
			\boldsymbol{\theta}^{(n+1)} &\gets (1-\eta)\boldsymbol{\theta}^{(n)} + \eta \boldsymbol{\theta}^{(n+1)} \\
			\boldsymbol{r}^{(n+1)} &\gets (1-\eta)\boldsymbol{r}^{(n)} + \eta \boldsymbol{r}^{(n+1)}
		\end{align*}
		\textbf{Compute} rate improvement $\Delta R = |R(\boldsymbol{\theta}^{(n+1)}, \boldsymbol{r}^{(n+1)}) - R(\boldsymbol{\theta}^{(n)}, \boldsymbol{r}^{(n)})|$; \\
		\textbf{Set} $n \gets n+1$;
	}
\end{algorithm}

\subsection{Convergence Behavior of Algorithm}
In this context, the FAS active port positions and the passive beamforming RIS phase-shift matrix at the $n$-th iteration, represented by $f\big( {\boldsymbol{r}}^{\left(n\right)};\boldsymbol{\theta}^{\left(n\right)}\big)$, with a given $\delta$. Therefore, we can observe the following:
\begin{equation}\label{eq.20}
 f\big({\boldsymbol{r}^{\left(n\right)}}; \boldsymbol{\theta}^{\left(n\right)}\big) {\geq} f\big({\boldsymbol{r}}^{\left(n+1\right)}; \boldsymbol{\theta}^{\left(n\right)}\big)  
\end{equation}
where (\ref{eq.20}) is valid because (\ref{eq.17}) represents a convex function. The positions of the FA ${\boldsymbol{r}}^{\left(n+1\right)}$ are optimized and adjusted during the $(n+1)^{\text{th}}$ iteration based on the passive beamforming $\boldsymbol{\theta}^{\left(n\right)}$. Notably, ${\boldsymbol{r}}^{\left(n+1\right)}$ remains consistent throughout the $(n+1)^{\text{th}}$ iteration, even after updating the passive beamforming to $\boldsymbol{\theta}^{\left(n+1\right)}$, thereby affirming the validity of (\ref{eq.20}). With regard to the constraint (\ref{eq.17}c), the sequence of objectives ensures $f\big( {\boldsymbol{r}};\boldsymbol{\theta}\big) \geq -\delta N$, signifying the convergence of the objective function $f\big( {\boldsymbol{r}}^{\left(n\right)};\boldsymbol{\theta}^{\left(n\right)}\big)$. Acknowledging the system's limited resources and the minimum SINR requirement, we posit that the objective function in (\ref{eq.17}) must have an identifiable lower bound, a finite value. As a result, we can establish the convergence of Algorithm \ref{algorithm1} towards a feasible solution.

\subsection{Computational Complexity Analysis}

Suppose we consider a scenario with a single UAV. It can be straightforwardly shown that there exists a total of $2(K+N+1)+1$ optimization variables in (\ref{eq.17}), assuming they are real-valued. Additionally, a total of $N+2$ second-order conic constraints are in place. Consequently, by relying on the rationale provided in \cite{7}, the comprehensive per-iteration complexity of the proposed SOCP-based approach is determined as:
\begin{equation}
\mathcal{O} [2(4+N)^{0.5} (1+K+N) (4+16K+8N+20K^2+8KL+4N^2)]
\end{equation}
In practical settings, the number of RIS meta-atoms is expected to be significantly larger than the count of interfering base stations. As a result, the complexity of Algorithm \ref{algorithm1} based on SCA can be reasonably estimated as $\mathcal{O}(\rm{N}^{3.5})$. Although the AO approach presented in \cite{23} simplifies the optimization process, the inherent coupling between design variables may limit the optimality of the solution. Moreover, the algorithm in \cite{23} requires a higher number of iterations to converge, leading to extended problem-solving durations.  

The mobility and physical constraints of UAVs introduce two main challenges for FAS deployment: mechanical instability (\textit{e.g.}, sway and vibration) and weight limitations, especially for lightweight aerial platforms. These issues, however, can be effectively mitigated. From a stability perspective, UAVs are commonly equipped with flight stabilizers, gimbals, and vibration isolation systems that reduce rapid motion \cite{p11}, enabling reliable FAS operation. Furthermore, FAS port switching operates over quasi-static channel coherence intervals \cite{p14}, meaning port decisions do not need to be real-time and can be robust against minor motion-induced perturbations. Regarding weight, recent advances in microfluidic and reconfigurable antenna hardware have demonstrated highly compact and lightweight FAS designs using liquid metal or programmable surfaces \cite{p15}. Such implementations are well-suited for integration into modern lightweight UAVs without significantly affecting payload or flight endurance. Therefore, both motion-related and weight-related concerns can be effectively addressed using current technology and careful system design.
\begin{figure}[!t]
\centering
\centerline{\includegraphics[width=0.52\textwidth]{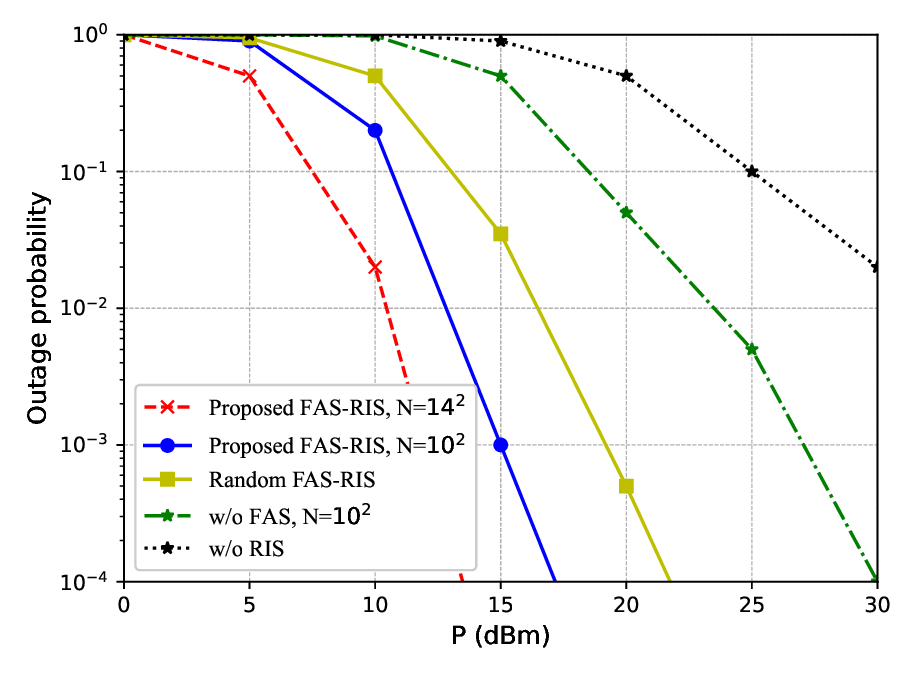}}
\caption{ Outage probability versus transmit power under CSI-free scheme.}
\label{fig.7}
\end{figure}
\section{Numerical Results}\label{sec.5}

This section presents simulation results to evaluate the performance of Algorithm \ref{algorithm1}. We consider a system with an RIS-assisted UAV, with the BS for the UAV, $\mathrm{BS_{UAV}}$ and $\mathrm{UAV_{RIS}}$ are located at $(0, 0, 10)$ m and $(x_{\text{U}}, y_{\text{U}}, 50)$ m, respectively. The other co-channel base stations are positioned at a height of 10 m, with their locations randomly and uniformly generated within their respective cells, designated by coordinates $(x_{\text{BS},k}, y_{\text{BS},k}, 10)$ m. The UAV operates at a fixed altitude of $h_\text{U}=50$ m and moves at a maximum speed of $V_{\text{max}}= 25$ m/s over a flight duration of $80$ seconds. In the simulation results, each BS is equipped with $N_t = 4$ antennas, and the UAV is configured with $M = 20$ ports. The channel is characterized by $L_t = L_r = 3$ multipath components for both the transmit and receive links. The number of co-channel interfering BSs is set to $K = 6$. The system is allocated a total bandwidth of $10$ MHz, operating at a center frequency of $5$ GHz. The parameters are set to $\delta = 0.001$ and $\sigma^2= -174$ dBm. The BS-UAV and UAV-$\mathrm{UAV_{RIS}}$ channels are modeled using Rician fading, characterized by a Rician factor $\mathbf{\bar{K}}=5$ and distance-dependent path loss for the communication links, as described in \cite{19}. Furthermore, we assume LoS-only links between the BSs and the $\mathrm{UAV_{RIS}}$, and between the $\mathrm{UAV_{RIS}}$ and the target UAV. The distance between the UAV and the BS is assumed to be a random variable that follows a uniform distribution, ranging from $50$ to $200$ m.
Algorithm \ref{algorithm1} is designated as the Proposed FAS, and its performance is compared against the following benchmark schemes: 
\begin{enumerate}
    \item \textbf{Random FAS-RIS}: In this configuration, the active ports of the UAV-mounted fluid antenna are selected randomly, and the phase shifts of the RIS meta-atoms are also assigned randomly without optimization.
    \item \textbf{FPA}: The UAV and the BS are equipped with four fixed-position antennas.
    \item \textbf{Traditional-AS}: In this case, the target UAV is equipped with conventional fixed-position antennas. A total of $n_s = 4$ antennas is activated and spaced by $\frac{\lambda}{2}$, while the phase shifts of the RIS meta-atoms are optimized.
\end{enumerate}

Fig.~\ref{fig.7} illustrates the relationship between outage probability and transmit power for various system configurations, with active ports $m_l = 6$. The performance of the proposed FAS-RIS system is compared against scenarios employing FAS without RIS, systems without FAS, and systems with random FAS-RIS. The results demonstrate the superior performance of the FAS-RIS configuration, highlighting the significant reduction in outage probability achieved through the integration of RIS. The integration of FAS and RIS further enhances performance.

\begin{figure}[!t]
\centerline{\includegraphics[width=0.51\textwidth]{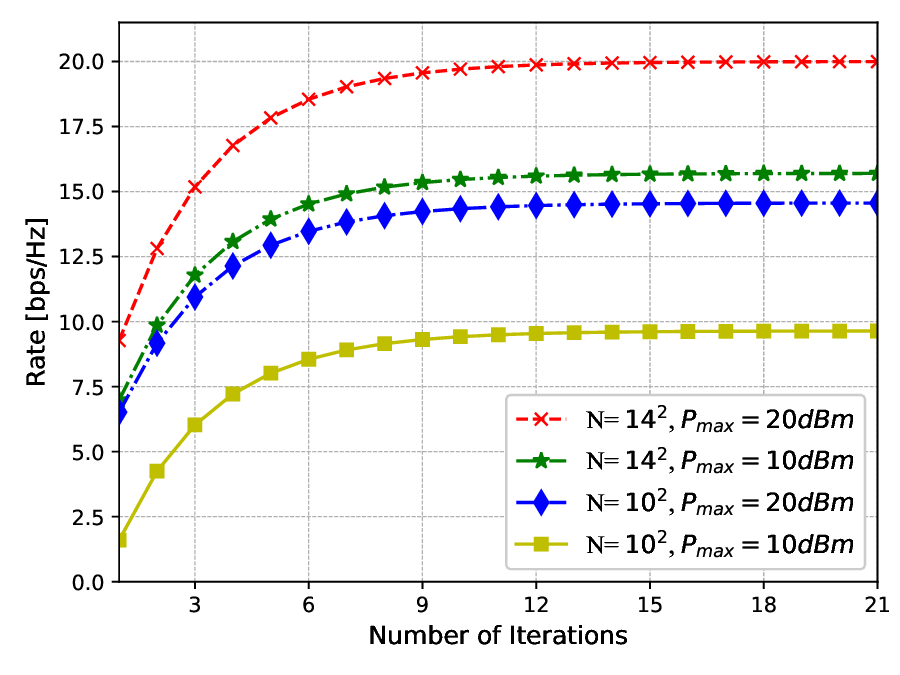}}
\centering
\caption{Convergence behavior of SCA-Algorithm.}
\label{fig.5}
\end{figure}

The convergence behavior of Algorithm~\ref{algorithm1} is illustrated in Fig.~\ref{fig.5} for two different values of $P_{\mathrm{max}}$. As shown in the figure, the proposed algorithm exhibits fast convergence, with the sum-rate stabilizing within approximately 10 iterations once a suitable penalty weight is identified that satisfies the optimization constraints. Furthermore, the results indicate that increasing the number of RIS meta-atoms leads to a higher achievable sum-rate. Additionally, it is also observed that a larger value of $P_{\mathrm{max}}$ corresponds to an increase in the required transmit power. In other words, improving the QoS target for the UAV demands greater power expenditure at the BS.

\begin{figure}[!t]
\centerline{\includegraphics[width=0.51\textwidth]{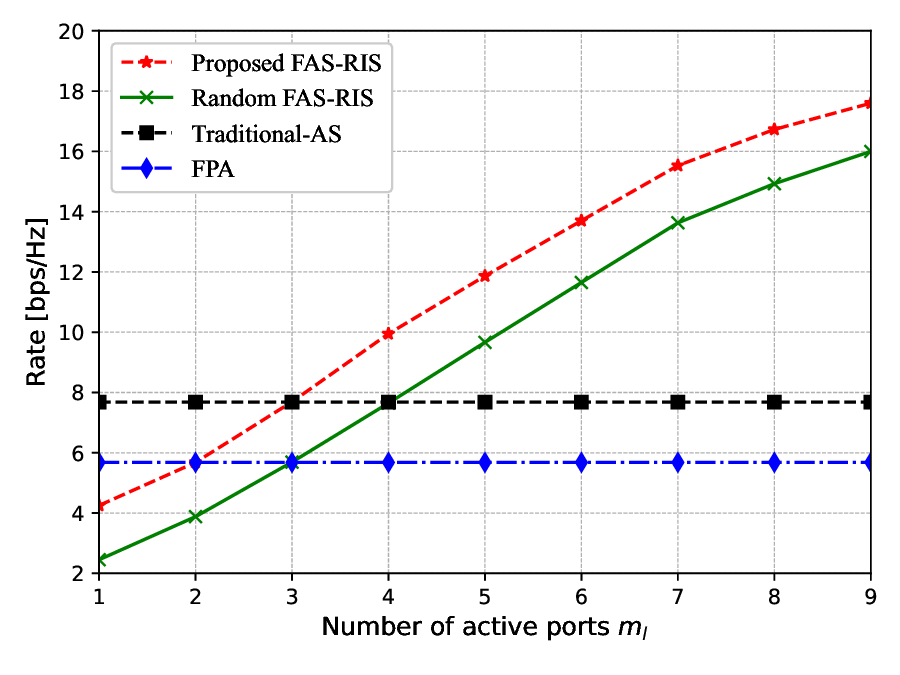}}
\centering
\caption{Performance comparison based on the number of activated ports.}
\label{fig.2}
\end{figure} 

Fig.~\ref{fig.2} presents the achievable rate $R$ performance of four different schemes as a function of the number of activated ports, $m_l$, with $P_{\mathrm{max}}=10$ dBm, and $N=~10^2$. The proposed FAS-RIS algorithm demonstrates a significant and consistent improvement in achievable rate as $m_l$ increases, owing to its ability to dynamically exploit spatial diversity. The Random FAS-RIS scheme also benefits from additional active ports but performs suboptimally due to the absence of coordinated optimization. In contrast, both the Traditional-AS and FPA baselines show flat performance across all values of $m_l$, as their antenna configurations do not adapt to spatial variations. Notably, when $m_l = 4$, the proposed FAS-RIS scheme achieves approximately $2.1$ dB performance gains over the Random FAS-RIS and Traditional-AS schemes, and $4.1$ dB higher than the FPA scheme. These results validate the advantage of integrating optimized FAS and RIS in enhancing link quality and spectral efficiency under UAV-mounted configurations.

\begin{figure}[!t]
\centerline{\includegraphics[width=0.51\textwidth]{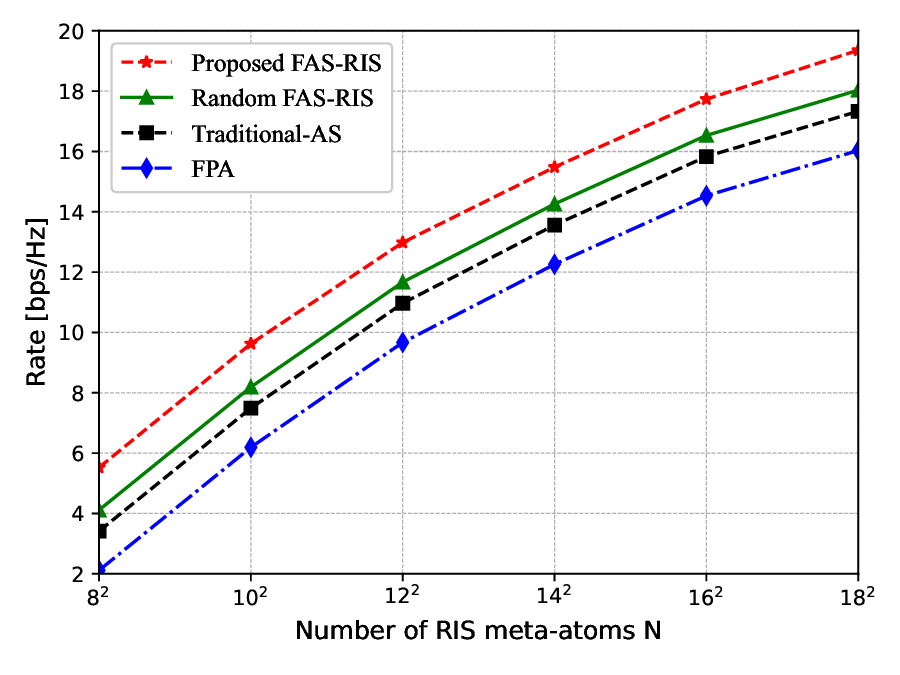}}
\caption{Achievable rate at the UAV versus the number of RIS meta-atoms.}
\label{fig.3}
\end{figure} 

\begin{figure}[!t]
\centerline{\includegraphics[width=0.510\textwidth]{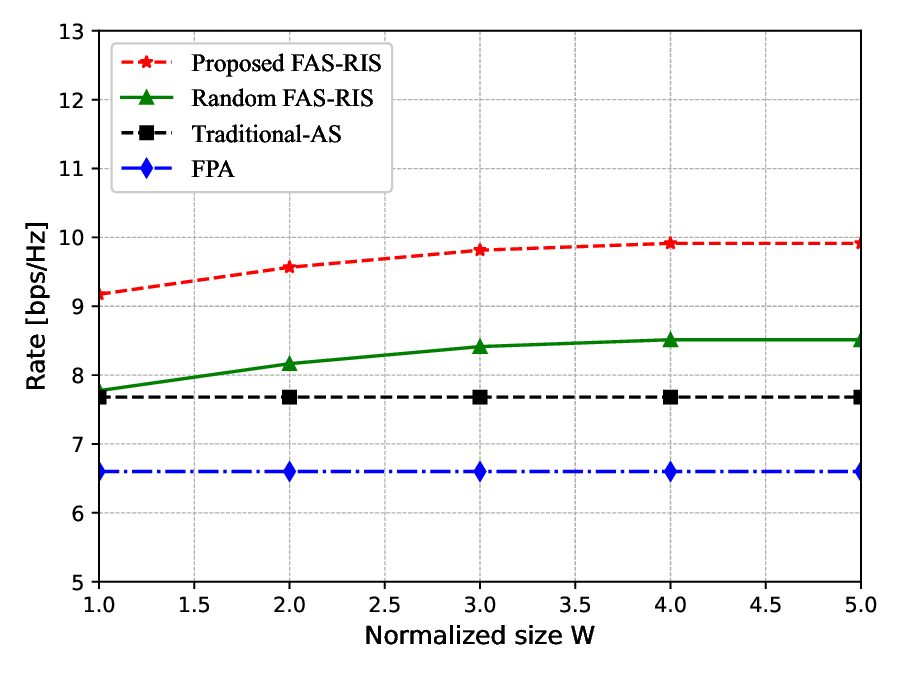}}
\caption{Achievable rate at the UAV versus the normalized sizes $W$ of FAS, where $N=10^2$, $P_{\mathrm{max}}=10$ dBm, and $m_l=4$.}
\label{fig.4}
\end{figure} 

Fig.~\ref{fig.3} shows the achievable rate at the UAV versus the number of RIS meta-atoms $N$. The number of activated ports $m_l$ is set to 4 in this figure. As $N$ increases, all schemes benefit from increased passive reflection area, which is most effectively exploited by the Proposed FAS-RIS due to its joint optimization. The Random FAS-RIS scheme, while not optimizing phases, still benefits from the increased element count $N$ as the passive reflection gain increases. The proposed FAS-RIS scheme consistently outperforms the baselines due to its joint optimization of RIS and FAS configuration. In contrast, Random FAS-RIS, Traditional-AS, and FPA offer lower gains due to limited adaptability. The performance gap increases with $N$, highlighting the efficiency of the proposed method in leveraging RIS capabilities for interference mitigation. The gap between the proposed method and the baselines widens with increasing $N$, confirming the effectiveness of the proposed joint SCA-SOCP algorithm in fully leveraging the RIS architecture and FAS flexibility for UAV communication scenarios.

Fig.~\ref{fig.4} illustrates the achievable rate $R$ at the UAV for different FAS schemes as the normalized size $W$ increases. The proposed optimized FAS-RIS consistently outperforms all baselines, achieving rates from approximately $7$ to $8$ bps/Hz as $W$ increases from $1$ to $5$, demonstrating the effectiveness of joint FAS active ports' position and RIS phase shifts optimization. As $W$ increases, the achievable rate improves due to the enhanced spatial degrees of freedom, allowing better signal reception and interference mitigation. However, beyond $W = 4$, the rate gain saturates, indicating diminishing returns. These results validate the advantages of FAS-RIS integration in mitigating interference and enhancing UAV communication performance, making it a promising approach for interference-limited environments.

\begin{figure}[!t]
\centerline{\includegraphics[width=0.510\textwidth]{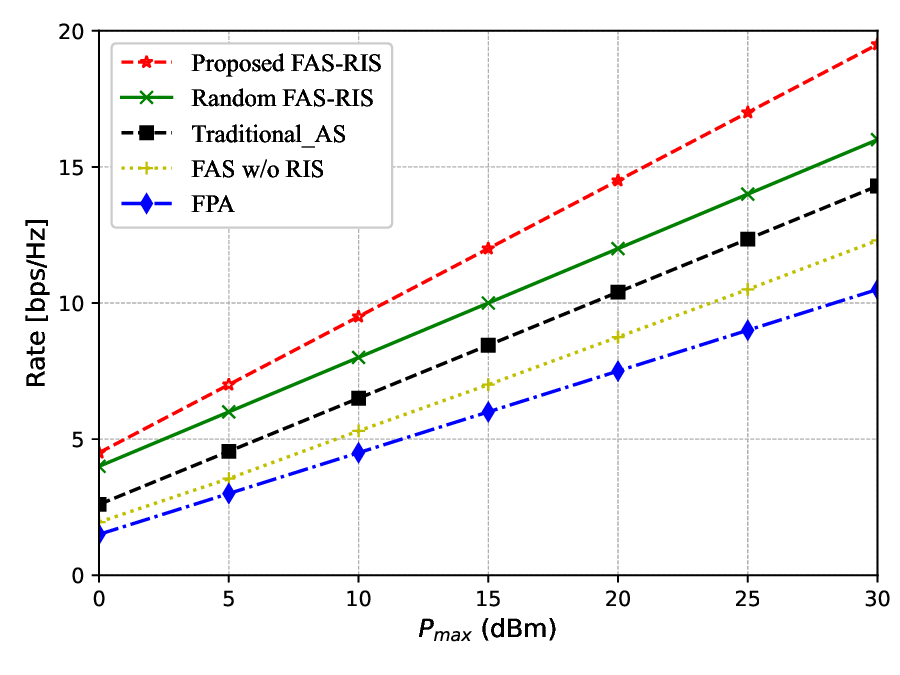}}
\caption{Achievable rate at the UAV versus the transmitted power $P_{\mathrm{max}}$, where $N=10^2$, $m_l=4$.}
\label{fig.6}
\end{figure}

Finally, Fig.~\ref{fig.6} illustrates the impact of the maximum transmit power $P_{\max}$ on the achievable rate $R$ at the target UAV. As expected, increasing $P_{\max}$ leads to a higher rate for all schemes due to enhanced signal strength. The proposed FAS-RIS scheme consistently achieves the highest performance across the entire power range, validating its effectiveness in interference suppression and signal enhancement. The FAS w/o RIS scheme, where only the fluid antenna is optimized, consistently underperforms relative to the full FAS-RIS scheme, underscoring the critical role of RIS optimization in further boosting link quality and mitigating interference. Moreover, the near-linear rate growth observed in the FAS-based schemes with increasing $P_{\max}$ indicates efficient power utilization, making the proposed solution particularly attractive for power-constrained UAV communication scenarios.
\section{Conclusion}\label{sec.6}

In this paper, we investigated the joint optimization of passive RIS phase shifts and FAS positions using a framework that incorporates the SCA algorithm and SOCP relaxation. This approach effectively tackles the challenge of maximizing the rate of UAV downlink communication while ensuring QoS. Importantly, the algorithm updates all optimization variables simultaneously in each iteration, which promotes efficient convergence. The integration of SOCP and SCA enables rapid convergence to a viable solution. Our results demonstrate that the FAS-RIS aided UAV downlink system achieves a significant rate of improvement compared to FPA-based systems. This enhancement is attributed to the superior spatial diversity and interference mitigation achieved by the joint optimization of FAS port positions and RIS elements in FAS-RIS, leading to higher achievable rates and lower outage probabilities. These results highlight the benefits of integrating the mobility of reflective elements into RIS design. In future work, we will investigate the effects of UAV altitude and speed on interference mitigation performance, as well as address the challenges posed by dynamic channel variations and imperfect CSI. Additionally, this study provides a framework for jointly optimizing UAV-mounted RIS and FAS in interference-aware communication systems. Future extensions will consider multi-UAV scenarios, where cooperative beam management and coordinated RIS–FAS operation can enable scalable, high-capacity aerial networks through distributed optimization and UAV-swarm coordination.

\section*{Acknowledgment}
The work of T. A. Tsiftsis was supported by the project NEURONAS. The research project NEURONAS is implemented in the framework of H.F.R.I call “3rd Call for H.F.R.I.’s Research Projects to Support Faculty Members and Researchers” (H.F.R.I. Project Number: 25726). The work of C.-B. Chae was in part supported by the Korean government (RS-2024-00428780, RS-2024-00428780). 

\bibliographystyle{IEEEtran}
\bibliography{ref}

\end{document}